%% file: main.tex
\documentclass[twocolumn]{article} 
\input{preamble} 
\begin{document}

\input{title} 
\maketitle

\input{01-introduction}
\input{02-related}
\input{03-method}
\input{04-case-study}
\input{05-results}
\input{06-discussion}

\clearpage
\appendix
\onecolumn 
\input{appendix}
\twocolumn

\clearpage
\bibliographystyle{plainnat}  
\bibliography{references}

\end{document}

%% file: preamble.tex
\usepackage{amsmath, amssymb}
\usepackage{graphicx}
\usepackage{microtype} 
\usepackage{caption} 
\usepackage{booktabs} 
\usepackage{subcaption} 
\usepackage{xurl}  
\usepackage{placeins}
\usepackage{tabularx}
\usepackage[numbers]{natbib} 
\usepackage{hyperref} 

\hypersetup{
  pdfauthor={Torsten Tiltack},
  pdftitle={AIJIM: A Scalable Model for Real-Time AI in Environmental Journalism},
  pdfsubject={Environmental Journalism, AI, XAI, Crowdsourcing},
  pdfkeywords={AIJIM, Environmental Journalism, Real-Time AI, GPT-4, NamicGreen, SHAP, LIME, XAI},
  colorlinks=true,
  linkcolor=blue,       
  citecolor=blue,       
  urlcolor=blue,        
  breaklinks=true,      
  pdfborder={0 0 0}     
}

%% file: title.tex
\title{AIJIM: A Scalable Model for Real-Time AI in Environmental Journalism}
\author{Torsten Tiltack}
\date{} 
\maketitle

\begin{abstract}
    This paper introduces AIJIM, the Artificial Intelligence Journalism Integration Model---a novel framework for integrating real-time AI into environmental journalism. AIJIM combines Vision Transformer-based hazard detection, crowdsourced validation with 252 validators, and automated reporting within a scalable, modular architecture. A dual-layer explainability approach ensures ethical transparency through fast CAM-based visual overlays and optional LIME-based box-level interpretations. Validated in a 2024 pilot on the island of Mallorca using the NamicGreen platform, AIJIM achieved 85.4\% detection accuracy and 89.7\% agreement with expert annotations, while reducing reporting latency by 40\%. Unlike conventional approaches such as Data-Driven Journalism or AI Fact-Checking, AIJIM provides a transferable model for participatory, community-driven environmental reporting, advancing journalism, artificial intelligence, and sustainability in alignment with the UN Sustainable Development Goals and the EU AI Act.
\end{abstract}


%% file: 01-introduction.tex
\sloppy
\section{Introduction}
\label{sec:introduction}

Environmental journalism plays a \textbf{critical role} in raising awareness and shaping policy on issues such as climate change, pollution, and biodiversity loss, directly influencing public discourse and evidence-based decision-making \cite{IPCC2023,Marshall2025}. Yet, traditional methods suffer from slow data acquisition, costly verification, and limited scalability—particularly in \textbf{under-monitored regions} where urgent environmental threats (e.g., oil spills, illegal deforestation) require rapid response \cite{Ekeh2025}.

\textbf{Artificial Intelligence (AI)}—including Natural Language Processing (NLP), Computer Vision (CV), and Machine Learning (ML)—has enhanced journalism through automation and large-scale analysis. However, existing approaches lack real-time adaptability in dynamic environmental contexts.

To address these gaps, we introduce the \textbf{Artificial Intelligence Journalism Integration Model (AIJIM)}: a \textbf{scalable, modular, and transferable model} for integrating AI into real-time environmental journalism. AIJIM combines citizen-generated imagery, explainable AI (XAI), and crowdsourced validation to ensure precise detection, low latency, and transparent accountability. Its architecture—leveraging Vision Transformers and adaptable NLP pipelines—ensures future flexibility and cross-domain applicability.

In a 2024 pilot study on Mallorca, AIJIM demonstrated its effectiveness within the NamicGreen platform by analyzing 1,000 geotagged images, detecting 50 undocumented waste sites with 85.4\% accuracy, and achieving 89.7\% expert agreement—while reducing reporting latency by 40\%.

Unlike existing models (e.g., Data-Driven Journalism, Computational Journalism), AIJIM bridges real-time detection, human-in-the-loop verification, and transparent reporting. This paper presents AIJIM’s conceptual foundation (Section~\ref{sec:AIJIM}), its practical implementation—demonstrated through detection overlays and heatmaps (Figs. 6–10)—and its implications for participatory, AI-driven journalism, while supporting the UN Sustainable Development Goals and compliance with the EU AI Act.

%% file: 02-related.tex
\section{Related Work}
\label{sec:background}

Existing research highlights the increasing integration of \textbf{Artificial Intelligence (AI)} in journalism, yet significant limitations persist. Traditional AI approaches, such as \textbf{Data-Driven Journalism (DDJ)}, depend primarily on structured, retrospective datasets, rendering them ineffective for real-time environmental reporting \cite{Ivancsics2019,Tong2022}. Similarly, \textbf{Computational Journalism}, which employs machine learning for complex pattern recognition, struggles with the dynamic and unstructured nature of live environmental data \cite{Fernandes2023,Korostin2025}. In addition, \textbf{AI Fact-Checking} models, which focus on textual claim verification, lack capabilities for real-time visual or spatial hazard detection \cite{Wolfe2024,Cazzamatta2025}.

Building on these insights, it is essential to evaluate how existing AI-driven journalism frameworks address these challenges. This section identifies their methodological constraints and positions the \textbf{Artificial Intelligence Journalism Integration Model (AIJIM)} as a novel approach to integrating \textbf{real-time automation, adaptability, and scalable journalistic workflows}.

\subsection{AI in Journalism}
AI technologies have significantly advanced journalism by enabling automation, rapid analysis, and streamlined verification processes. \textbf{Natural Language Processing (NLP)} has enhanced automated news generation from structured datasets, thereby improving journalistic efficiency \cite{OpenAI2023,Tong2022}. Additionally, developments in \textbf{Computer Vision (CV)} have facilitated rapid detection and classification of visual data, surpassing traditional manual methods in both speed and accuracy \cite{Tiltack2025,Floridi2021}. Despite these advancements, current AI-driven methods often rely heavily on periodic updates from structured sources, such as satellite imagery, limiting their effectiveness in dynamic and rapidly evolving environmental scenarios. This gap underscores the need for AI solutions like AIJIM, which are capable of handling dynamic, real-time data streams efficiently.

\subsection{Challenges in Environmental Journalism}
Environmental journalism frequently faces practical challenges, especially with regard to the timely acquisition and verification of reliable data. This is particularly evident in \textbf{remote or under-monitored regions}, complicating rapid responses to urgent environmental threats \cite{Marshall2025}. Conventional journalism methods—characterized by manual data collection and expert-driven verification—are resource-intensive, prone to inaccuracies, and hinder both scalability and timely intervention \cite{Ivancsics2019,deLimaSantos2022}. Consequently, there is an urgent demand for innovative AI-driven solutions to improve the precision, timeliness, and overall credibility of environmental reporting.

\subsection{Limitations of Existing AI Models}
Most AI models for environmental monitoring rely on \textbf{satellite-based imagery} and long-term trend analysis, which are effective for \textbf{large-scale environmental assessments} but insufficient for detecting \textbf{localized, high-frequency hazards} such as illegal dumping or micro-scale pollution \cite{Ivancsics2019,Tong2022}. Additionally, existing AI solutions often \textbf{fail to adjust adequately for contextual variability}, resulting in misclassifications in underrepresented geographic regions \cite{deLimaSantos2022}. Regulatory constraints—including \textbf{GDPR compliance and data privacy risks}—further limit the scalability of these models \cite{Floridi2021}.

To address these gaps, AI models for environmental journalism must integrate \textbf{real-time data processing, multi-source validation, and scalable automation}. The AIJIM model explicitly tackles these shortcomings by providing a modular, adaptable \textbf{theoretical structure} for real-time hazard detection and automated reporting.

\subsection{Comparison of AIJIM with Existing Journalism Models}
Although several AI-based journalism models have been proposed, none fully meet the specific demands of real-time environmental crisis reporting. \textbf{Data-Driven Journalism (DDJ)}, focused on historical structured datasets, lacks immediate adaptability \cite{Tong2022}. Similarly, \textbf{Computational Journalism}, while proficient at analyzing extensive datasets, struggles with dynamic, real-time environmental scenarios \cite{Sallami2024}. Meanwhile, \textbf{AI Fact-Checking} excels primarily in textual verification but falls short in processing real-time visual and spatial data, which are critical for environmental monitoring \cite{Cazzamatta2025}.

AIJIM addresses these limitations by uniquely integrating real-time hazard detection, adaptive community validation, and automated reporting into a coherent, modular framework. This combination ensures \textbf{scalability, adaptability, and ethical AI governance}, bridging the gap between automation and human oversight in environmental journalism. A detailed methodological comparison is presented in Section \ref{sec:AIJIM}.

Unlike purely technical frameworks or platform-specific toolkits, AIJIM is deliberately conceptualized as a \textbf{theoretical model}. Its goal is to offer a transferable structure that enables systematic analysis and design of AI-supported environmental journalism workflows. By integrating technological, ethical, and participatory dimensions, AIJIM contributes to ongoing efforts in theory-building at the intersection of journalism studies, human-centered AI, and digital environmental communication.

While model-agnostic explainability approaches like LIME and SHAP are well established in AI research, their practical application in real-time object detection remains limited due to performance constraints and architectural mismatch. In contrast, activation-based methods such as CAM offer high interpretability for convolutional models by visualizing spatial attention, and are therefore widely used in visual explanation pipelines. AIJIM combines both approaches by using CAM overlays as a fast default explainability layer and providing optional LIME-based per-box interpretation for expert validation.

%% file: 03-method.tex
\section{The Artificial Intelligence Journalism Integration Model (AIJIM)}
\label{sec:AIJIM}

The \textbf{Artificial Intelligence Journalism Integration Model (AIJIM)} advances beyond conventional AI-driven journalism by integrating real-time hazard detection, crowdsourced validation, and automated reporting into a scalable, modular, and adaptable framework. Unlike traditional models, AIJIM is designed for dynamic environmental journalism, enabling flexible integration of AI technologies tailored to specific reporting needs.

AIJIM’s modular architecture supports multiple AI components, ensuring adaptability. It integrates Vision Transformer-based DETR models, alternative CV and NLP architectures, and crowdsourced validation modules, allowing independent adjustment without system redesign~\cite{Tiltack2025}. This modularity enables AIJIM to address diverse environmental crises, from localized pollution to large-scale disasters, with rapid deployment across geographical regions.

AIJIM’s scalability supports high-volume data processing while maintaining accuracy and ethical responsibility. In the 2024 Mallorca pilot, AIJIM processed 1,000 citizen-generated images in hours, demonstrating efficiency without compromising quality~\cite{Tiltack2025}. Its adaptability integrates emerging technologies like satellite imagery or IoT sensors, enhancing capabilities in urban and remote areas.

Beyond detection, AIJIM enhances credibility through crowdsourced validation, achieving an 89.7\% agreement rate with expert annotations in the Mallorca pilot~\cite{Tiltack2025}. For reporting, AIJIM employs adaptable NLP models, reducing latency by 40\%~\cite{OpenAI2023}. By converging automation and human oversight, AIJIM delivers scalable, accurate, and ethically responsible journalism, applicable across investigative and large-scale monitoring contexts.

\subsection{Methodology}
\label{subsec:methodology}
AIJIM’s effectiveness was evaluated using a Sequential Explanatory Mixed-Methods Approach~\cite{Creswell2020}, combining quantitative and qualitative analyses:
\begin{itemize}
    \item \textbf{Quantitative Analysis:} The 2024 Mallorca pilot tested AIJIM’s real-time capabilities, analyzing 1,000 geotagged citizen images to detect 50 undocumented waste sites~\cite{Tiltack2025,McGovern2022}. The NamicGreen platform served as a proof-of-concept, with the DETR-based Vision Transformer enabling robust detection, though AIJIM supports alternative models.
    \item \textbf{Qualitative Analysis:} Twelve expert interviews with journalism, AI, and regulatory professionals informed the evaluation. Structured Qualitative Content Analysis (QCA) confirmed thematic dimensions with a Krippendorff’s Alpha of 85\%, ensuring reliability~\cite{Mayring2019}.
\end{itemize}

\subsection{Technology Adaptability}
\label{subsec:tech_adaptability}
AIJIM’s architecture supports diverse AI models and hardware environments. Its modular structure enables integration of alternative object detection and NLP models based on deployment needs, ensuring scalability and flexibility. As a technology-agnostic model, AIJIM adopts future AI advancements without structural changes, optimizing performance across journalistic contexts~\cite{Tiltack2025}.

\subsection{Comparison with Existing AI-Based Journalism Models}
\label{subsec:comparison_ddj}
AIJIM surpasses traditional AI-driven journalism models—such as Data-Driven Journalism (DDJ), Computational Journalism, AI Fact-Checking, and Satellite AI—by enabling real-time, multi-modal environmental reporting. Unlike DDJ, which relies on retrospective datasets, AIJIM processes unstructured, citizen-generated data instantly, providing immediate environmental insights. AIJIM's modular design allows the integration of various AI models for enhanced adaptability, addressing limitations in existing frameworks. A detailed comparison across key dimensions is presented in Table~\ref{tab:AIJIM_vs_models} and Table~\ref{tab:ai_journalism_comparison} in Appendix~\ref{app:tables}.

\begin{itemize}
    \item \textbf{DDJ} relies on structured, retrospective datasets, limiting real-time responsiveness~\cite{Tong2022,Stalph2023}. AIJIM processes unstructured, citizen-generated images instantly.
    \item \textbf{Computational Journalism} excels in large-scale data analysis but struggles with dynamic environmental data~\cite{Cools2024,Nedungadi2025}. AIJIM’s real-time detection and validation address this.
    \item \textbf{AI Fact-Checking} verifies textual claims but lacks visual hazard detection~\cite{Dierickx2024}. AIJIM integrates multi-modal analysis with crowdsourcing.
    \item \textbf{Satellite AI} supports large-scale monitoring but is constrained by low resolution and periodic updates~\cite{Dai2024,Gondwe2024}. AIJIM’s ground-level imagery ensures granularity and immediacy.
\end{itemize}

\begin{figure}[htbp]
    \centering
    \includegraphics[width=0.5\textwidth]{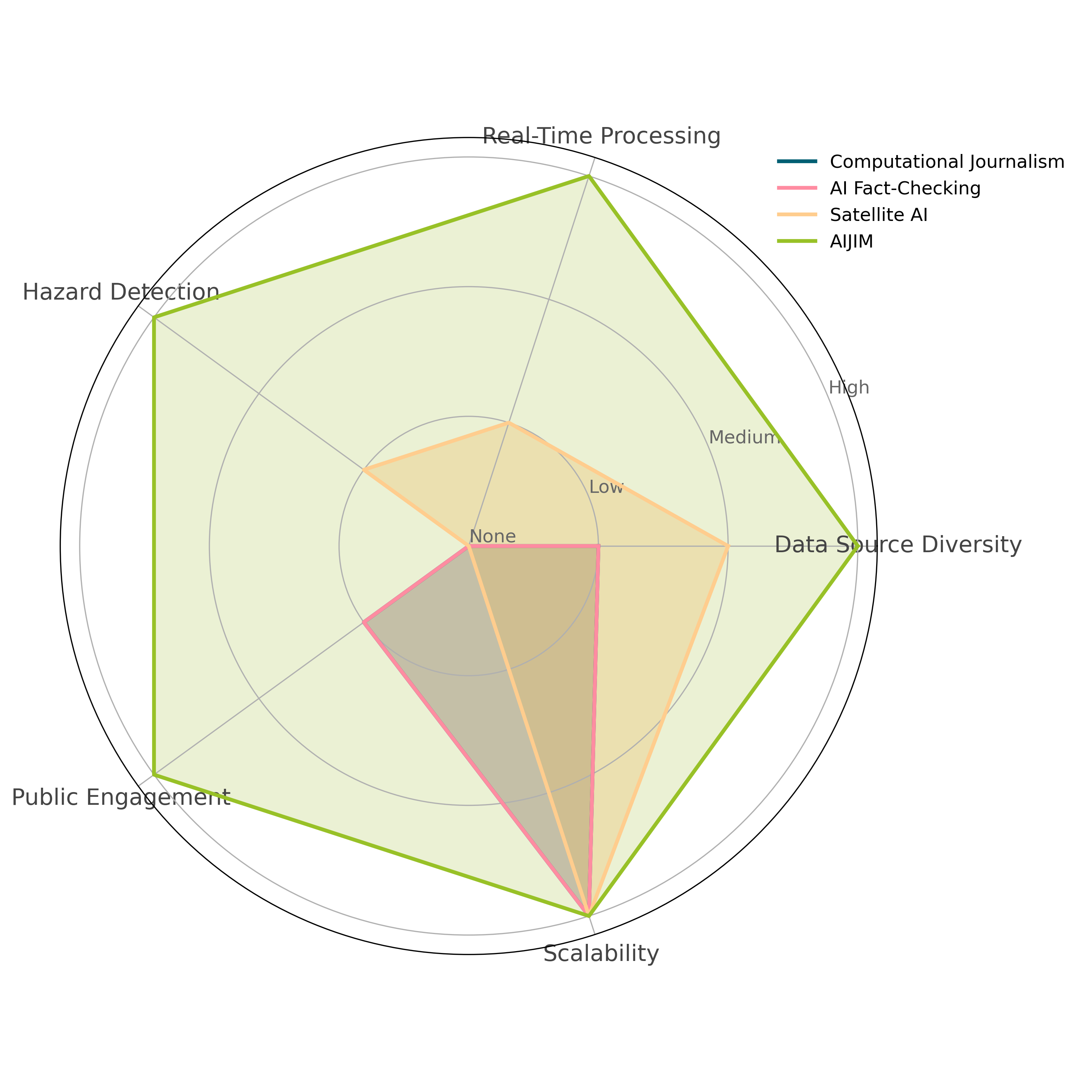}
    \caption{Radar chart comparing AIJIM with established AI journalism models across data source diversity, real-time processing, hazard detection, and public engagement. AIJIM excels in critical dimensions for environmental reporting.}
    \label{fig:radar_ai_comparison}
\end{figure}

As shown in Fig.~\ref{fig:radar_ai_comparison}, AIJIM surpasses existing models in real-time processing, public engagement (252 validators), and small-scale hazard detection, supported by Explainable AI (XAI) for transparency~\cite{Shin2021}. The 2024 Mallorca pilot validated these strengths, achieving 40\% latency reduction and 85.4\% accuracy~\cite{Tiltack2025}, establishing a new benchmark for scalable, participatory journalism.

\subsection{AIJIM Workflow}
\begin{figure}[htbp]
    \centering
    \includegraphics[width=0.5\textwidth]{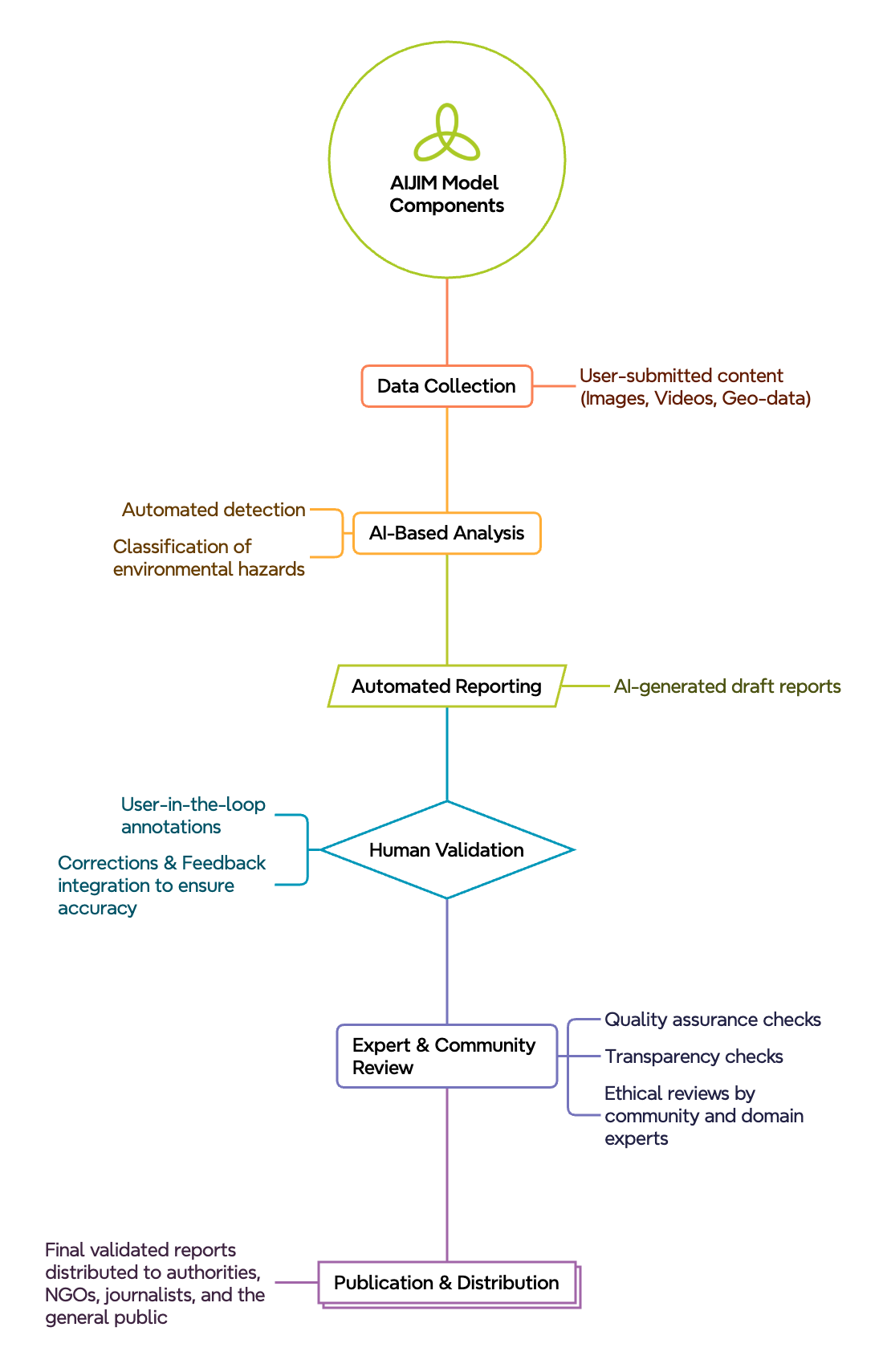} 
    \caption{AIJIM’s six components: (1) Data Collection; (2) AI-Based Analysis; (3) Automated Reporting; (4) Human Validation; (5) Expert \& Community Review; (6) Publication \& Distribution.}
    \label{fig:AIJIM_model}
\end{figure}

AIJIM follows a six-step validation and reporting process, as illustrated in Fig.~\ref{fig:AIJIM_model}:
\begin{enumerate}
    \item \textbf{Data Collection:} Users upload geotagged images via platforms like NamicGreen.
    \item \textbf{AI-Based Analysis:} A Vision Transformer-based DETR model enables real-time detection, with support for alternative models~\cite{Tiltack2025}.
    \item \textbf{Automated Reporting:} Adaptable NLP models generate draft reports.
    \item \textbf{Human Validation:} Crowdsourced validators verify detections through image annotations, ensuring accuracy.
    \item \textbf{Expert \& Community Review:} Experts and community members validate reports for ethical compliance.
    \item \textbf{Publication \& Distribution:} Validated reports are published and distributed to stakeholders.
\end{enumerate}

\subsection{Core Components}
\subsubsection{Automated Environmental Monitoring}
\label{subsec:image_recognition}
AIJIM employs a DETR-based Vision Transformer, optimized with ONNX and TensorRT, for real-time hazard detection~\cite{Tiltack2025}. Its modularity supports models like YOLO or Swin Transformer, chosen for:
\begin{itemize}
    \item Superior detection of small, obscured hazards.
    \item Lower computational cost than YOLOv8 or Faster R-CNN.
    \item Robustness across lighting conditions~\cite{Carion2020,Tiltack2025}.
\end{itemize}
The 2024 Mallorca pilot analyzed 1,000 images, achieving 85.4\% ± 1.8\% accuracy and identifying 50 waste sites~\cite{Tiltack2025,Li2024}.

\subsection{Stakeholder Interaction and Validation}
\label{subsec:validation}
AIJIM overlays hazards onto images for visual validation by 252 community validators, achieving 89.7\% expert agreement in the Mallorca pilot~\cite{Tiltack2025}. Interactive maps and gamification enhance engagement~\cite{Verma2024}. Expert interviews emphasized:
\begin{itemize}
    \item Efficiency requires XAI transparency~\cite{GarciaTapiaMateo2025}.
    \item Slight rural performance variation needs dataset expansion~\cite{Floridi2023}.
    \item Hybrid AI-human collaboration ensures integrity~\cite{Verma2024}.
    \item Human-in-the-loop validation aligns with EU AI Act 2025~\cite{Gstrein2024,Bolda2024}.
\end{itemize}

\subsection{AI-Based Narrative Generation}
\label{subsec:AIJIM_narrative}
AIJIM’s reporting module generates narratives, reducing latency by 40\%~\cite{Tiltack2025}. Supporting NLP models like GPT-4, Claude, Gemini, and LLaMA, it adapts to linguistic and ethical needs~\cite{OpenAI2023,Anthropic2024,Google2024}. The Mallorca pilot validated object classification (e.g., plastic foil), producing actionable reports (Appendix, Abbildung~\ref{app:fig:AIJIM_object_detection},~\ref{app:fig:AIJIM_report}). Features include:
\begin{itemize}
    \item Dynamic summarization for tailored reports.
    \item Context-aware tone adjustment.
    \item GIS visualization via Mapbox APIs~\cite{Tiltack2025}.
\end{itemize}
XAI techniques (SHAP) ensure transparency, mitigating bias, as confirmed by experts~\cite{Floridi2023,Verma2024}.

\subsection{Empirical Advantages of AIJIM}
AIJIM outperforms traditional methods, with the Mallorca pilot showing:
\begin{itemize}
    \item 40\% latency reduction via real-time inference~\cite{Tiltack2025}.
    \item 89.7\% validation agreement~\cite{Tiltack2025}.
    \item Expert-backed credibility~\cite{Pareek2024}.
\end{itemize}

\subsection{Performance Comparison with Traditional Methods}
AIJIM’s improvements in efficiency, accuracy, and engagement are detailed in Section~\ref{sec:results}, Table~\ref{tab:AIJIM_performance}.

\subsection{Explainability Architecture: CAM and LIME}
\label{sec:explainability}

AIJIM’s dual-layer explainability uses Class Activation Mapping (CAM), as illustrated in Appendix~\ref{app:cam_example}, for real-time heatmaps and optional LIME for detailed audits. CAM visualizes detection focus, while LIME perturbs regions for deeper insight, balancing scalability and interpretability~\cite{Tiltack2025}.

LIME provides an additional layer of transparency by highlighting the specific regions of an image that contribute to the model’s decision. A detailed example of this process for a "metal can" detection can be found in Appendix~\ref{app:lime_example}. The image in the appendix illustrates the areas of interest identified by LIME and their impact on the classification decision, offering a deeper understanding of the model’s behavior.

%% file: 04-case-study.tex
\section{Implementation in NamicGreen}
\label{sec:implementation}

The \textbf{Artificial Intelligence Journalism Integration Model (AIJIM)} was implemented in \textit{NamicGreen}, an AI-powered platform for real-time environmental monitoring and reporting. This proof-of-concept deployment showcased AIJIM’s ability to process high-volume citizen-generated data with scalability and efficiency.

\textbf{Technical Implementation:} NamicGreen integrates AI-driven components for hazard detection, validation, and reporting, as detailed in Section~\ref{sec:AIJIM}. Its modular design supports cloud-based processing for large-scale analyses and edge AI deployment for remote areas with limited connectivity, ensuring real-world applicability~\cite{Tiltack2025}. The platform’s flexible reporting workflow adapts to various NLP models, extending beyond the GPT-4 implementation used in this study~\cite{OpenAI2023}.

NamicGreen exemplifies AIJIM’s balance of automation and human oversight, leveraging crowdsourced validation and \textbf{Explainable AI (XAI)} to enhance accuracy and transparency, as validated in the 2024 Mallorca pilot (Section~\ref{subsec:mallorca-pilot}).

\subsection{System Architecture}
\label{subsec:architecture}

NamicGreen integrates seven core components, with a focus on scalable design and real-time data processing. Details on specific subsystems such as AI image recognition and crowdsourced validation can be found in Sections~\ref{subsec:image_recognition} and~\ref{subsec:validation}.
\begin{itemize}
    \item \textbf{User/NamicGreen Drone}: Captures geotagged images and GPS data.
    \item \textbf{AI Analysis \& Classification}: Performs real-time pollution detection, adaptable to various CV models (see Section~\ref{subsec:image_recognition}).
    \item \textbf{Geolocation \& GIS Integration}: Maps data via Flask API for geospatial insights.
    \item \textbf{Automated Report Generation}: Produces draft reports using NLP models (see Section~\ref{subsec:AIJIM_narrative}).
    \item \textbf{Community Validation}: Refines outputs through crowdsourced feedback (see Section~\ref{subsec:validation}).
    \item \textbf{Final Report \& Data Export}: Delivers validated reports to stakeholders.
    \item \textbf{AI Feedback Loop}: Continuously improves model performance.
\end{itemize}

\begin{figure}[htbp]
    \centering
    \includegraphics[width=0.5\textwidth]{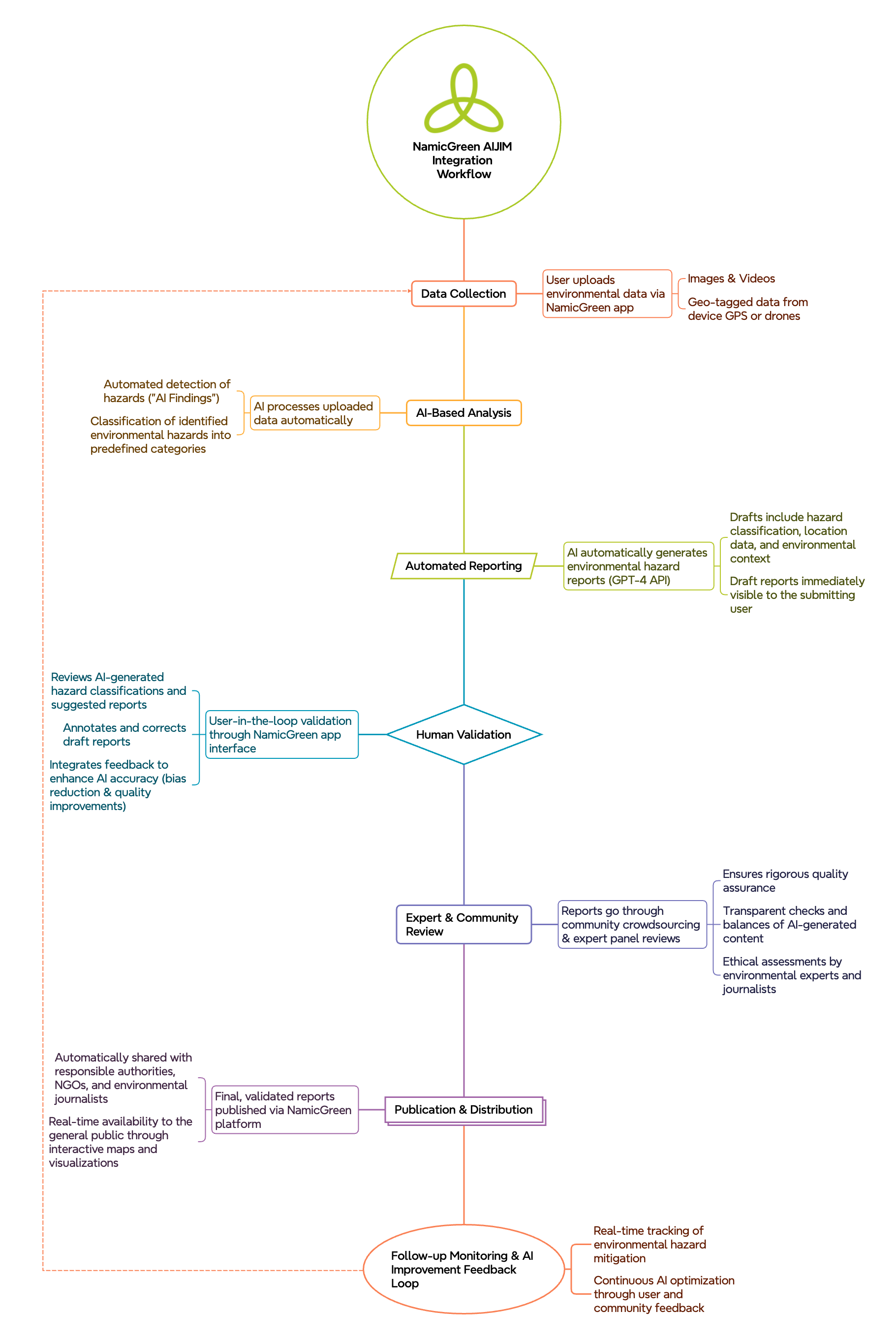}
    \caption{NamicGreen workflow: (1) Data Collection; (2) AI-Based Analysis; (3) Automated Reporting; (4) Human Validation; (5) Expert \& Community Review; (6) Publication \& Distribution; (7) AI Feedback Loop. Arrows depict data flows and iterative feedback.}
    \label{fig:architecture}
\end{figure}

\subsection{AI Image Recognition Module}
\label{subsec:ai-image-recognition}

NamicGreen employs advanced Computer Vision models for rapid hazard detection, as described in Section~\ref{subsec:image_recognition}. Trained on diverse geotagged imagery, the system ensures accurate detection, with ambiguous cases escalated for crowdsourced validation~\cite{Tiltack2025}.

\subsection{Crowdsourced Validation}
\label{subsec:crowdsourced-validation}

NamicGreen’s crowdsourced validation system enhances accuracy using a weighted voting mechanism, as detailed in Section~\ref{subsec:validation}. Validators prioritize high-confidence predictions, with expert review for ambiguous cases, ensuring reliability~\cite{Tiltack2025}.

\subsection{Data Visualization and Automated Reporting}
\label{subsec:visualization-reporting}

NamicGreen transforms data into actionable insights using GIS-based heatmaps powered by Mapbox APIs, pinpointing pollution hotspots for journalists and policymakers. Automated reporting, detailed in Section~\ref{subsec:AIJIM_narrative}, leverages NLP models to produce rapid, validated reports~\cite{Tiltack2025}.

\subsection{Real-World Deployment: Mallorca Pilot Study}
\label{subsec:mallorca-pilot}

The 2024 Mallorca pilot validated AIJIM’s potential, processing 1,000 citizen-submitted geotagged images to uncover 50 undocumented illegal waste sites. Key outcomes include:
\begin{itemize}
    \item \textbf{Latency Reduction:} 40\% faster reporting cycle.
    \item \textbf{Community Engagement:} 252 active validators.
    \item \textbf{Detection Accuracy:} 85.4\% ± 1.8\%, surpassing manual methods~\cite{Tiltack2025}.
\end{itemize}

AIJIM’s dual explainability module (Section~\ref{sec:explainability}) deployed CAM-based overlays for real-time validation and a LIME-based API for selective expert audits, balancing transparency and scalability. Future developments include scaling to 10,000 images, integrating multispectral satellite analysis, and expanding model diversity (e.g., YOLO, Claude) and multilingual capabilities~\cite{Ivancsics2019,Tiltack2025}.

%% file: 05-results.tex
\section{Results and Discussion}
\label{sec:results}

The evaluation of the \textbf{Artificial Intelligence Journalism Integration Model (AIJIM)} within the NamicGreen platform demonstrates its efficacy and scalability. The 2024 Mallorca pilot benchmarks AIJIM against traditional journalism, with expert insights highlighting technical and ethical considerations~\cite{Tiltack2025}.

\subsection{Performance Evaluation}
\label{subsec:performance_evaluation}

The 2024 Mallorca pilot processed 1,000 citizen-generated geotagged images, detecting 50 undocumented waste sites with an accuracy of 85.4\% $\pm$ 1.8\%. Crowdsourced validation by 252 participants achieved 89.7\% expert agreement, reducing reporting latency by 40\% compared to traditional methods~\cite{Tiltack2025}. A 5-fold cross-validation (95\% confidence interval, $p < 0.05$) ensured the robustness of these results. Detailed metrics are provided in Table~\ref{tab:AIJIM_performance} in Appendix~\ref{app:tables}.

The pilot utilized a Vision Transformer-based DETR model (Section~\ref{subsec:image_recognition}) for real-time hazard detection, while crowdsourced validation (Section~\ref{subsec:validation}) significantly contributed to both model accuracy and public engagement.

\begin{figure}[htbp]
    \centering
    \includegraphics[width=0.9\linewidth]{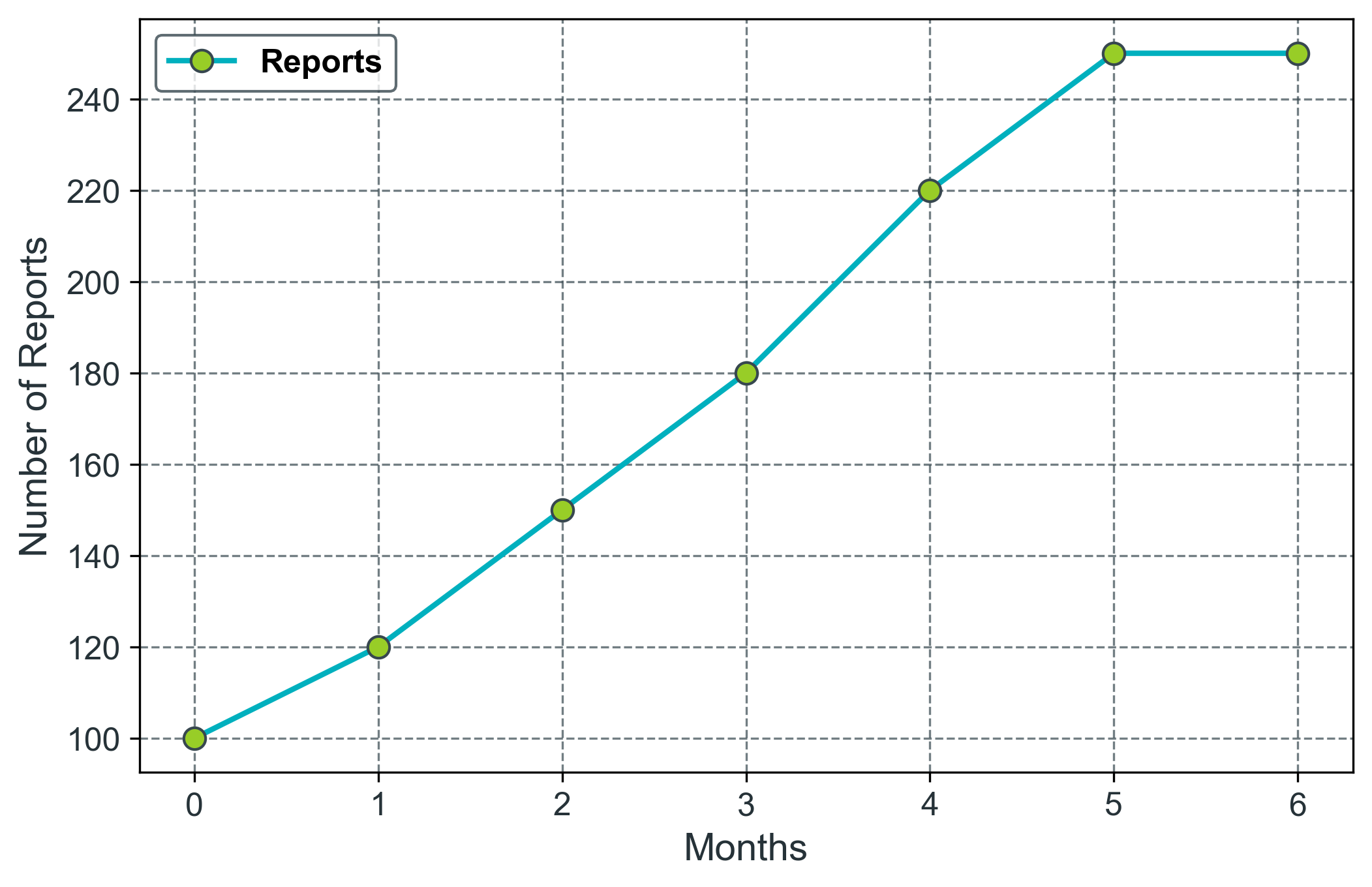}
    \caption{Increase in the number of reports submitted via the NamicGreen platform over six months, demonstrating rising public engagement and the adoption of AI-powered environmental reporting.}
    \label{fig:report_growth}
\end{figure}

Figure~\ref{fig:report_growth} illustrates the steady increase in the number of environmental reports submitted via the NamicGreen platform over six months. This trend highlights the growing adoption of AIJIM’s AI-powered hazard detection and reporting features. The continuous rise in user-generated reports suggests that real-time detection and crowdsourced validation foster ongoing public engagement, reinforcing AIJIM’s scalability and long-term usability.

\subsection{Stakeholder Perspectives}
\label{subsec:stakeholder_perspectives}

Qualitative insights from twelve expert interviews (journalism, AI, regulatory fields) are summarized in Table~\ref{tab:expert_interviews} in Appendix~\ref{app:tables}.

\paragraph{AI Accuracy and Geographic Bias}
Experts praised AIJIM’s precision but noted slight rural performance variations, requiring expanded datasets (Section~\ref{subsec:ai_bias})~\cite{Tiltack2025}.

\paragraph{Journalistic Trust}
Rapid reporting was valued, but XAI tools (SHAP, LIME) were critical for credibility~\cite{Shin2021}.

\paragraph{Public Engagement}
Crowdsourcing was effective, though misinformation safeguards were recommended~\cite{Verma2024}.

\paragraph{Regulatory Challenges}
GDPR compliance was confirmed, but EU AI Act 2025 accountability gaps need journalism-specific ethics~\cite{Gstrein2024}.

\subsection{Addressing Expert Concerns: Future Strategies}
Expert feedback drives AIJIM’s evolution:
\begin{itemize}
    \item \textbf{Dataset Expansion:} Enhance diversity for improved global accuracy.
    \item \textbf{Explainable AI (XAI):} Expand integration of transparency tools (e.g., SHAP, LIME).
    \item \textbf{Crowdsourced Validation:} Implement credibility-weighted scoring mechanisms.
    \item \textbf{Regulatory Compliance:} Develop clear guidelines aligned with journalism-specific AI ethics.
\end{itemize}

\subsection{Comparison with Traditional Methods}
\label{subsec:comparison_traditional}

AIJIM outperforms traditional journalism, as shown in Figure~\ref{fig:performance_comparison} and Table~\ref{tab:AIJIM_vs_traditional_comparison} in Appendix~\ref{app:tables}. Pilot results (Section~\ref{subsec:performance_evaluation}) highlight 40\% faster processing and fivefold participation~\cite{Broussard2019,Tiltack2025,Kazaz2021}.

AIJIM integrates established technologies like Vision Transformers (DETR), CAM, and LIME, but its true innovation lies in the \textbf{modular integration} of these models into a \textbf{real-time, scalable framework} designed specifically for environmental journalism. By leveraging \textbf{crowdsourced validation} and providing \textbf{explainable AI} (XAI), AIJIM goes beyond the capabilities of traditional journalism models, offering \textbf{greater transparency} and \textbf{higher engagement}.

While traditional methods rely on \textbf{retrospective data collection} and \textbf{manual verification}, AIJIM utilizes \textbf{real-time hazard detection}, allowing for \textbf{immediate reporting} with \textbf{improved accuracy} and \textbf{participation}. This combination of \textbf{existing AI technologies} with \textbf{innovative workflows} marks AIJIM as a breakthrough model in the field of automated environmental journalism.

\begin{figure}[htbp]
    \centering
    \includegraphics[width=0.55\textwidth]{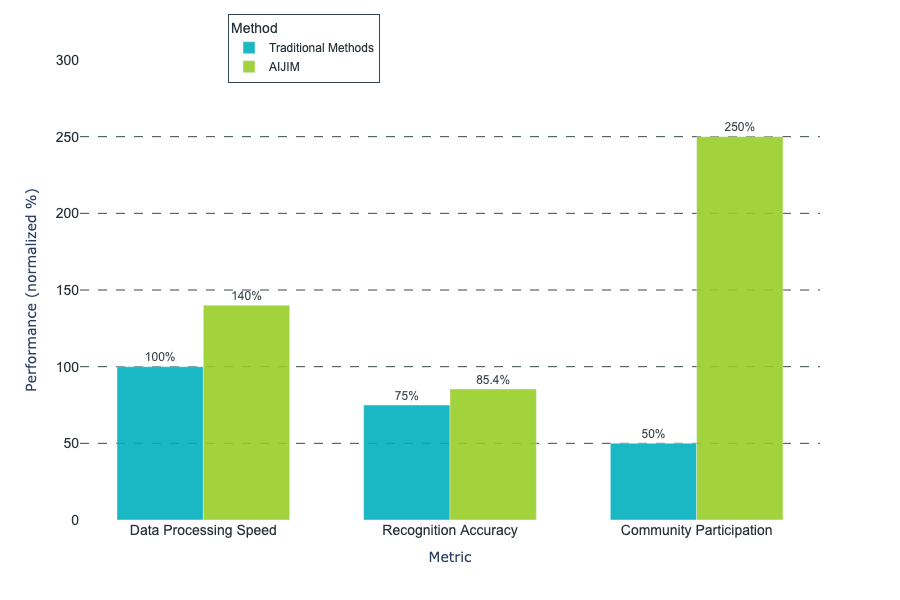}
    \caption{AIJIM vs. traditional journalism (Mallorca 2024): (1) 40\% faster processing, (2) 85.4\% accuracy, (3) 5x participation~\cite{Tiltack2025}.}
    \label{fig:performance_comparison}
\end{figure}

AIJIM’s real-time detection overcomes traditional journalism’s limitations by offering \textbf{faster reporting}, \textbf{increased scalability}, and \textbf{greater public involvement}.

\subsection{Challenges and Ethical Considerations}

AIJIM’s advancements come with \textbf{technical, ethical, and regulatory hurdles}, vital for its global rollout.

\subsubsection{AI Bias and Generalizability}
\label{subsec:ai_bias}

AIJIM’s 85.4\% accuracy shows slight rural variations due to limited training data, environmental complexity, and reduced crowdsourcing participation~\cite{McGovern2022,Tiltack2025,Gondwe2024}. Strategies include:
\begin{itemize}
    \item \textbf{Expanded Datasets:} Scale to 10,000 images~\cite{Tiltack2025}.
    \item \textbf{Adaptive Learning:} Use multispectral imagery~\cite{Li2024}.
    \item \textbf{Synthetic Data Augmentation:} Diversify exposure~\cite{McGovern2022}.
    \item \textbf{Geo-Weighted Validation:} Prioritize rural reports~\cite{Gondwe2024}.
\end{itemize}

\subsubsection{Misinformation Risks}
\label{subsec:misinformation_risks}

AI-generated reports risk false positives due to biased datasets or misinterpretation~\cite{AlZoubi2024,Cazzamatta2025}. Safeguards include uncertainty scoring and hybrid verification (Section~\ref{subsec:AIJIM_narrative})~\cite{Wolfe2024,Dierickx2024}.

\subsubsection{Transparency and Governance}
\label{subsec:transparency_governance}

AIJIM ensures transparency through XAI tools (SHAP, LIME) and GDPR-compliant encryption, aligning with EU AI Act 2025 (Section~\ref{subsec:validation}). Future governance will develop ethical guidelines for AI-driven journalism, incorporating federated learning to enhance accountability~\cite{Floridi2021,Gstrein2024,Tiltack2025}.

\subsubsection{Unintended Risks}
\label{subsec:unintended_risks}

Crowdsourcing mitigates bias but requires oversight to prevent misinformation~\cite{AlZoubi2024}.

\subsection{Explainability Outcomes}
\label{subsec:explainability_outcomes}

The Mallorca pilot study validated AIJIM's effectiveness in detecting 50 undocumented waste sites from 1,000 citizen-generated geotagged images. The model achieved 85.4\% accuracy, with crowdsourced validation contributing to an 89.7\% expert agreement rate. This demonstrates AIJIM's ability to process large datasets quickly, reducing reporting latency by 40\% compared to traditional methods. Furthermore, the use of Vision Transformer-based DETR models ensured rapid detection, and crowdsourced validation enhanced both accuracy and engagement. These results confirm AIJIM's potential as a scalable, real-time tool for environmental reporting.

%% file: 06-discussion.tex
\section{Conclusion and Future Work}
\label{sec:conclusion}

The \textbf{Artificial Intelligence Journalism Integration Model (AIJIM)} redefines environmental journalism by seamlessly integrating \textbf{real-time hazard detection}, \textit{AI-driven reporting (e.g., via GPT-4 in this study)}, and \textbf{crowdsourced validation}. The \textbf{2024 Mallorca pilot} on the NamicGreen platform showcased its prowess: achieving an \textbf{85.4\% detection accuracy} \textit{with a DETR-based model}, uncovering \textbf{50 previously undocumented illegal waste sites}, slashing reporting latency by \textbf{40\%} \textit{using GPT-4-generated reports}, and mobilizing \textbf{252 citizen validators} \cite{Tiltack2025}. \textit{These results, achieved with one possible configuration, mark a leap forward in \textbf{speed, precision, and engagement} over traditional methods (Table \ref{tab:AIJIM_performance}), positioning AIJIM as a pioneering, adaptable force in AI-driven journalism.}

By integrating \textbf{real-time citizen data} with \textbf{community-powered validation}, AIJIM enhances the efficiency and transparency of environmental journalism. This impact extends across both practical applications and scientific advancements.

\subsection{Practical and Scientific Implications}

\textbf{Practical Impact:} AIJIM empowers rapid hazard response and amplifies \textbf{public engagement}, transforming citizens from passive observers into active agents of change. By democratizing environmental journalism, AIJIM enables communities to take ownership of their local environmental issues, fostering accountability and collective action. By bridging the gap between data, technology, and community participation, AIJIM fosters a more informed, responsive, and participatory society in the fight against environmental crises.  
\textbf{Scientific Contribution:} It offers a \textbf{proven, adaptable model} for global journalism, aligning with the \textbf{United Nations Sustainable Development Goals (SDGs)} to bolster \textbf{climate resilience and sustainability}.

\subsection{Future Development Priorities}

Despite AIJIM’s proven advancements in AI-driven journalism, maintaining its long-term relevance demands ongoing innovation and refinement. As the global media landscape evolves, the following key priorities will drive AIJIM’s ability to remain at the forefront of AI-enhanced journalism:
\begin{itemize}
    \item \textbf{Dataset Expansion:} Scaling from \textbf{1,000 to 10,000 images} to enhance \textbf{geographic precision} and eliminate bias in underserved regions.
    \item \textbf{Advanced Remote Sensing:} Incorporating \textbf{multispectral imagery} for sharper detection in remote areas.
    \item \textbf{Federated Learning:} Pioneering \textbf{privacy-preserving AI training} to meet \textbf{GDPR standards}.
    \item \textbf{Ethical AI:} Embedding \textbf{SHAP-driven transparency} to uphold \textbf{journalistic integrity}.
    \item \textbf{Model Diversification:} \textit{Exploring alternative detection models (e.g., YOLO, Swin Transformer) and NLP systems (e.g., Claude, Gemini) to broaden AIJIM’s applicability across diverse journalistic and hardware contexts.}
\end{itemize}

\subsection{AI, Journalism, and Societal Impact}

AIJIM heralds a new era where AI amplifies journalism’s societal role, not by replacing human expertise, but by enhancing it. By fostering a symbiotic relationship between AI-driven automation and human judgment, AIJIM ensures that journalism remains both data-driven and deeply ethical. Future research must refine \textbf{AI-assisted editing}, strengthen \textbf{human-AI collaboration}, and ensure \textbf{transparent practices} to maintain \textbf{public trust} and accountability.

\subsection{International Regulatory Considerations}

AIJIM complies with the \textbf{EU AI Act 2025} and \textbf{GDPR} through \textbf{AES-256 encryption} and \textbf{data anonymization}. However, global adoption depends on adapting to evolving \textit{AI ethics standards and regulatory frameworks}.

\subsection{Conclusion}

AIJIM exemplifies AI’s \textbf{revolutionary potential} in environmental journalism, delivering a \textbf{real-time, scalable, and ethical} solution for global hazard detection. \textit{By explicitly integrating dual-level human validation—crowdsourced visual hazard verification and human review of AI-generated textual reports—with AI-driven insights, AIJIM ensures comprehensive accuracy and transparency. Demonstrated through the Mallorca pilot configuration, AIJIM does not just enhance journalism—it redefines its very foundation, shaping a future where technology and citizen collaboration drive impactful, evidence-based environmental action adaptable to various models and contexts} \cite{Tiltack2025}.

\subsection*{Explainability Trade-offs and Practical Implications}

AIJIM’s explainability strategy combines fast visual attribution through CAM overlays with the option of model-agnostic, per-box explanations using LIME. This dual-layer approach provides both efficiency for real-time applications and depth for expert-level validation. In practice, the CAM method supports immediate visual feedback with low computational cost, while LIME explanations—available via an additional API—enable in-depth audits and detailed traceability when needed. This balance ensures that AIJIM remains scalable without compromising interpretability in high-stakes scenarios.

\subsection*{Explainability Trade-offs and Practical Implications}

AIJIM’s explainability architecture balances real-time performance with interpretability through a two-layer approach. The default CAM-based visualization is computationally efficient and integrates directly into the detection pipeline. It has proven useful for general user feedback and field-level reporting.

On the other hand, the LIME module—though slower—is integrated as an optional, asynchronous feature for per-box inspection. It offers localized, model-agnostic insight that can support expert reviews and auditing processes. Together, these layers provide a practical trade-off between speed and explanatory depth, aligning well with AIJIM’s real-world requirements.

%% file: appendix.tex
\section{Appendix A: Detailed Tables}
\FloatBarrier
\label{app:tables}

\begin{table}[htbp]
\caption{Comparison of AIJIM with Other AI-Based Journalism Models}
\label{tab:AIJIM_vs_models}
\begin{tabularx}{\textwidth}{@{} l X X X X @{}}
\toprule
\textbf{Feature} & \textbf{DDJ} & \textbf{Comp. Journalism} & \textbf{AI Fact-Checking} & \textbf{AIJIM} \\
\midrule
Data Source & Structured datasets & Structured datasets & News archives & Real-time citizen images \\
Real-Time & No & No & No & Yes (0.1s latency) \\
Scalability & Limited & High & High & High (AI + crowdsourcing) \\
Bias Mitigation & Manual oversight & Algorithmic tweaks & Cross-checking & XAI (SHAP, audits) \\
Hazard Detection & None & Partial & None & Yes (85.4\% accuracy) \\
Engagement & Passive & Minimal & None & High (252 validators) \\
\bottomrule
\end{tabularx}
\end{table}

\begin{table}[htbp]
\caption{Comparison of AIJIM with Other AI Journalism Models}
\label{tab:ai_journalism_comparison}
\begin{tabularx}{\textwidth}{@{} l X X X X @{}}
\toprule
\textbf{Feature} & \textbf{Comp. Journalism} & \textbf{AI Fact-Check} & \textbf{Satellite AI} & \textbf{AIJIM} \\
\midrule
Data Source & Structured databases & News archives & Satellite imagery & Crowdsourced images + AI \\
Real-Time Processing & No & No & Limited & Yes \\
Hazard Detection & No & No & Limited (macro-scale) & Yes (85.4\% accuracy) \\
Public Engagement & Passive readers & Passive readers & No & High (252 validators) \\
Scalability & High & High & High & High \\
\bottomrule
\end{tabularx}
\end{table}

\begin{table}[htbp]
\caption{Performance Metrics of AIJIM (Mallorca Pilot Study, 2024)}
\label{tab:AIJIM_performance}
\begin{tabularx}{\textwidth}{@{} l X @{}}
\toprule
\textbf{Metric} & \textbf{Measured Value} \\
\midrule
Total analyzed images & 1,000 citizen-generated images \\
Illegal waste sites identified & 50 previously undocumented sites \\
Detection Accuracy (Precision - BoxP) & 85.4\% \\
Recall (R) & 59.7\% \\
Mean Average Precision (mAP@0.50) & 70.2\% \\
Mean Average Precision (mAP@0.50-0.95) & 55.9\% \\
Inference latency & 0.1 seconds per image (achieved with DETR, ONNX/TensorRT GPU in this implementation; adaptable to other models and hardware) \\
AI-Human Validation Agreement & 89.7\% \\
Active Community Validators & 252 participants \\
Reporting Latency Reduction & 40\% faster than traditional methods \\
Bias in Rural Detection Accuracy & Slight variation due to dataset and environmental factors, addressed through expanded datasets and adaptive learning \cite{Tiltack2025} \\
\bottomrule
\end{tabularx}
\end{table}

\begin{table}[htbp]
    \caption{Expert Perspectives on AIJIM’s Strengths and Challenges}
    \label{tab:expert_interviews}
    \begin{tabularx}{\textwidth}{@{} l X X @{}}
    \toprule
    \textbf{Theme} & \textbf{Strengths} & \textbf{Challenges} \\
    \midrule
    AI Accuracy & Precision (85.4\%) outpaces manual methods \cite{Tiltack2025} & Slight rural performance variation requires dataset diversity \cite{Tiltack2025} \\
    Trust & XAI (SHAP, LIME) boosts transparency \cite{Shin2021} & Black-box concerns demand clearer accountability \\
    Engagement & 252 validators enhance reliability and inclusion \cite{Verma2024} & Misinformation risks need robust controls \\
    Regulation & GDPR and EU AI Act alignment \cite{Gstrein2024} & Accountability for AI outputs remains unresolved \\
    \bottomrule
    \end{tabularx}
\end{table}

\begin{table}[htbp]
\caption{AIJIM vs. Traditional Journalism (Mallorca 2024)}
\label{tab:AIJIM_vs_traditional_comparison}
\begin{tabularx}{\textwidth}{@{} l X X @{}}
\toprule
\textbf{Feature} & \textbf{Traditional} & \textbf{AIJIM} \\
\midrule
Data Collection & Manual fieldwork & Crowdsourced AI \\
Verification & Expert-only & AI + 252 validators \\
Accuracy & Variable \cite{Kazaz2021} & 85.4\% \\
Latency & Hours/Days & 0.1s/image \\
Engagement & Low & High \\
Scalability & Limited & High \\
Bias & Subjective & XAI audits \\
Compliance & Manual & GDPR, EU AI Act \\
\bottomrule
\end{tabularx}
\end{table}

\clearpage
\section{Appendix B: Visual Examples of AIJIM Implementation}
\FloatBarrier
\label{app:visuals}

The following figures provide visual examples of AIJIM’s functionality, as demonstrated through its implementation in the NamicGreen platform, highlighting its real-world application in environmental journalism:

\begin{figure}[htbp]
    \centering
    \includegraphics[width=0.9\linewidth]{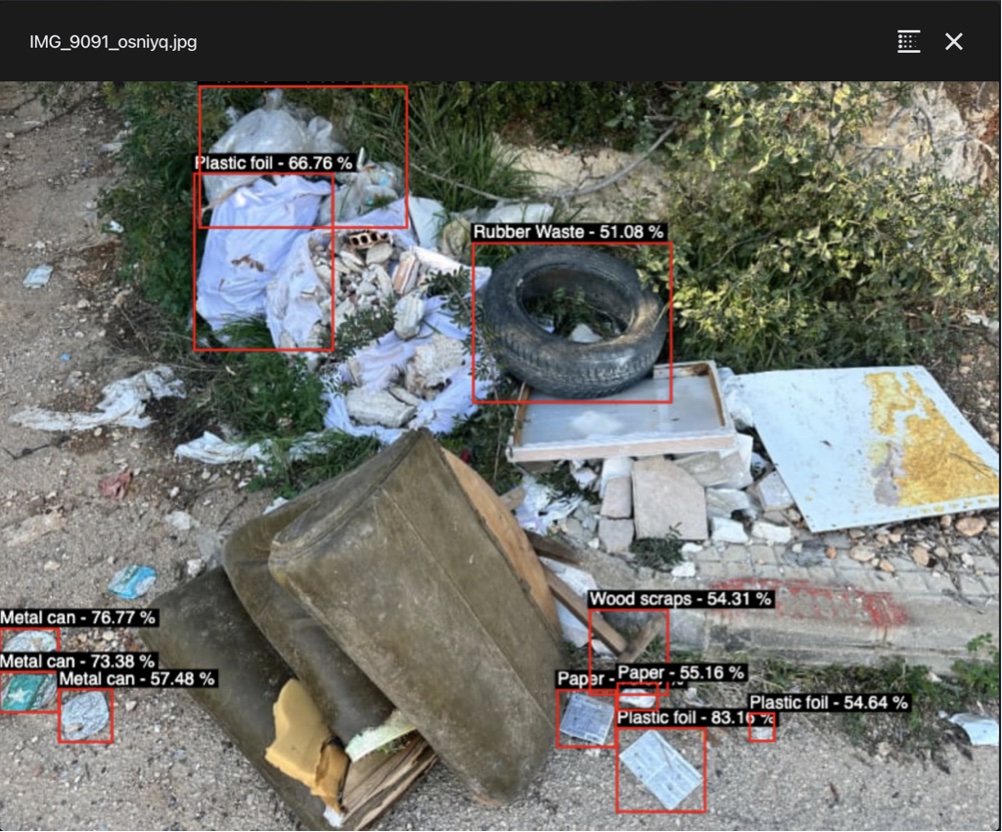}
    \caption{AIJIM Object Detection Implementation: Detailed view of AI-detected hazards (e.g., plastic foil, rubber waste) in user-submitted images, showcasing the model’s real-time analysis capabilities (referenced in Section \ref{subsec:AIJIM_narrative}).}
    \label{app:fig:AIJIM_object_detection}
\end{figure}

\begin{figure}[htbp]
    \centering
    \includegraphics[width=0.9\linewidth]{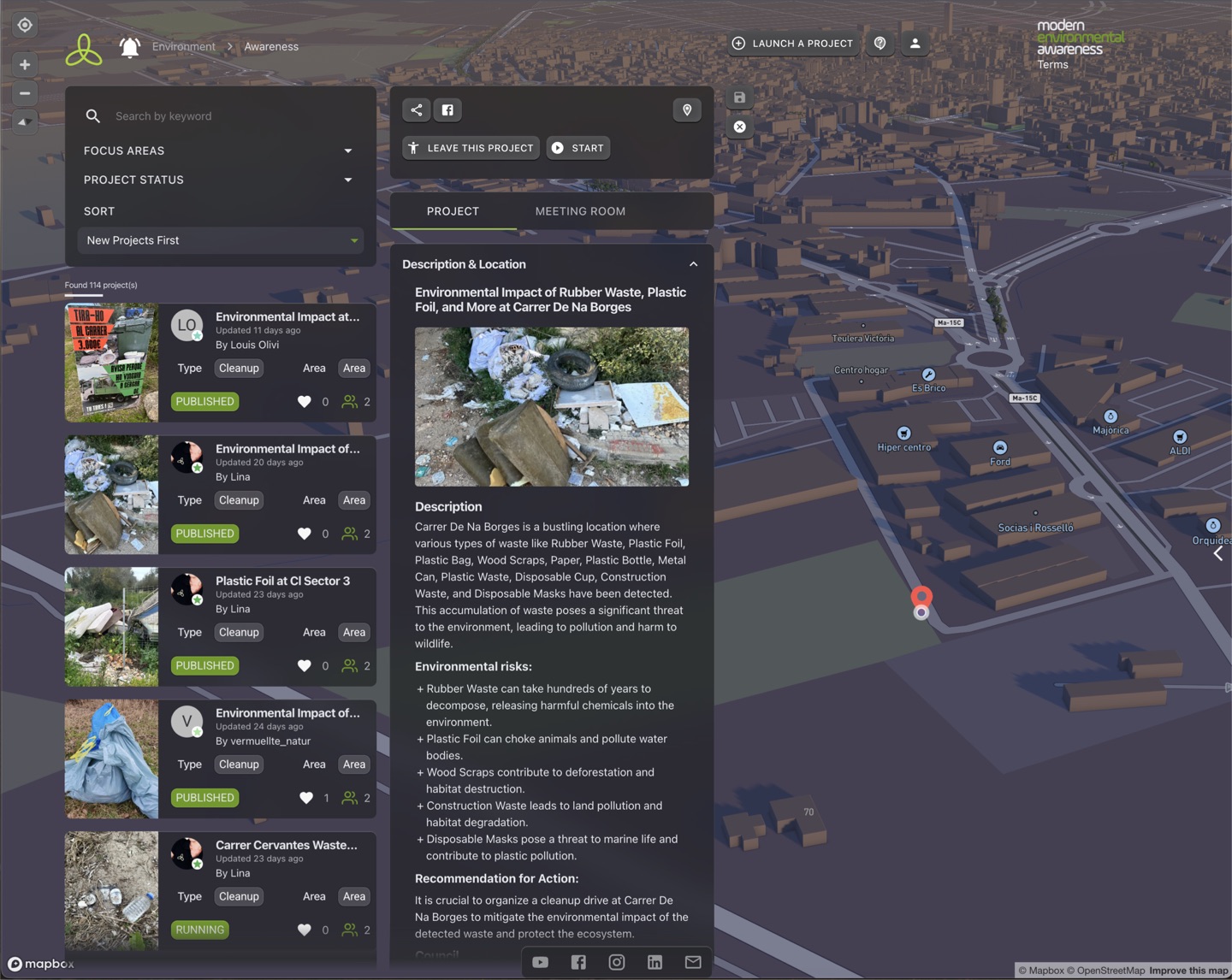}
    \caption{AIJIM Automated Report Generation: Example of an AI-generated environmental report based on object detection findings, demonstrating actionable insights (referenced in Section \ref{subsec:AIJIM_narrative}).}
    \label{app:fig:AIJIM_report}
\end{figure}

\begin{figure}[htbp]
    \centering
    \includegraphics[width=0.9\linewidth]{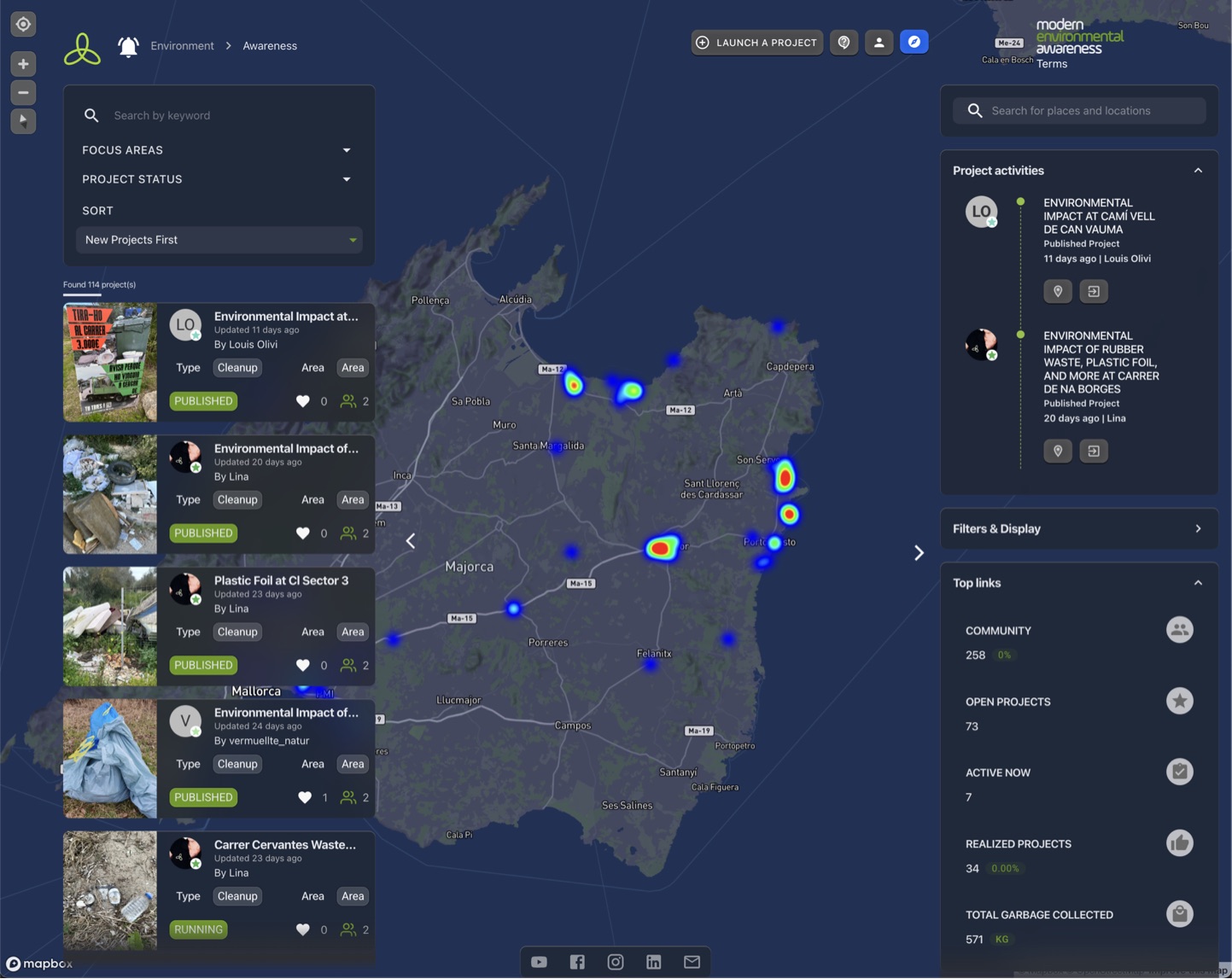}
    \caption{AIJIM Heatmap Visualization: Hotspots of environmental pollution in Mallorca, illustrating the model’s spatial data analysis capabilities (referenced in Section \ref{subsec:performance_evaluation}).}
    \label{app:fig:AIJIM_heatmap}
\end{figure}

\begin{figure}[htbp]
    \centering
    \includegraphics[width=0.9\linewidth]{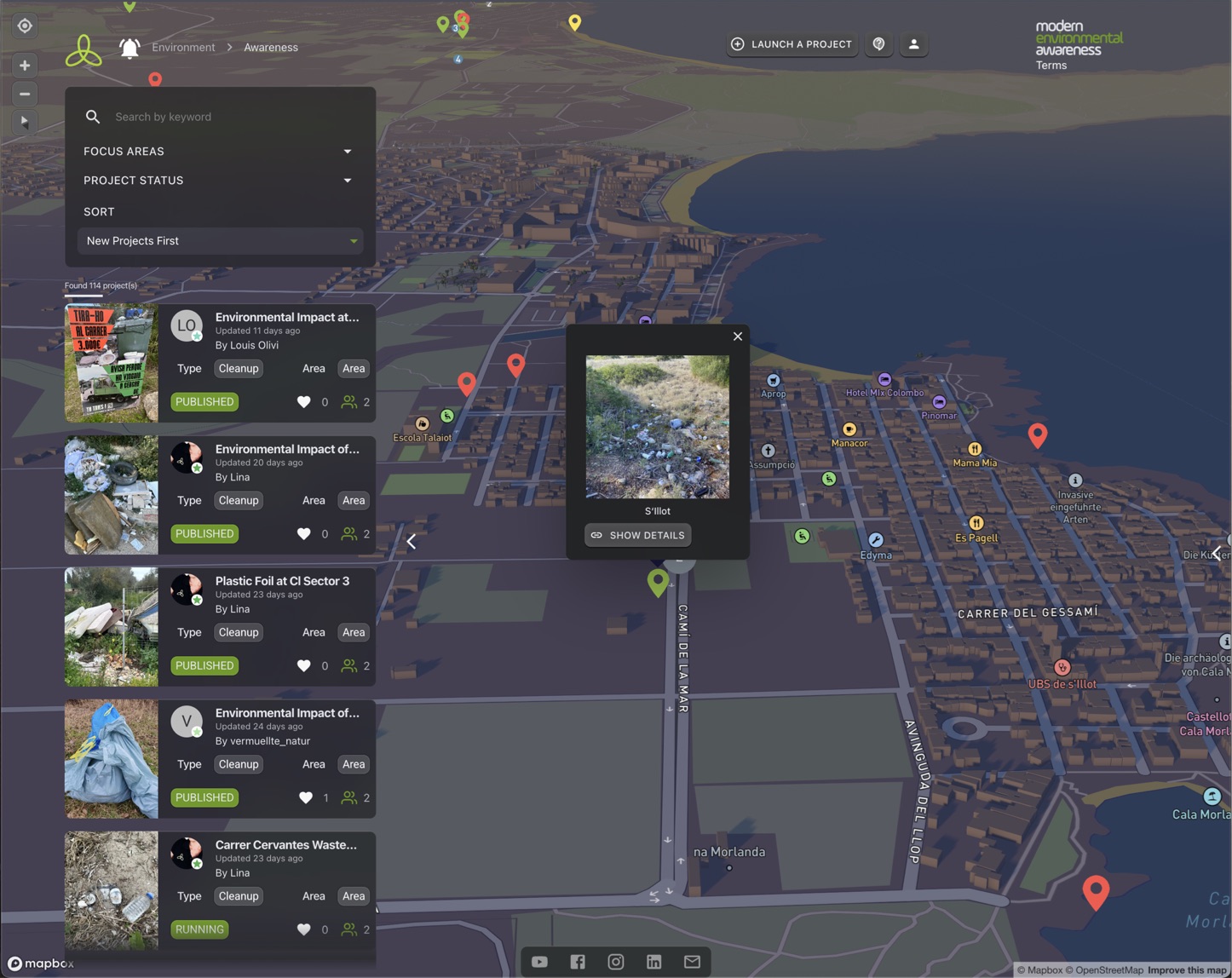}
    \caption{AIJIM Citizen Engagement Map: Overview of user-generated environmental reports, demonstrating the model’s participatory data collection (referenced in Section \ref{subsec:validation}).}
    \label{app:fig:AIJIM_map_reports}
\end{figure}

\begin{figure}[htbp]
    \centering
    \includegraphics[width=0.9\linewidth]{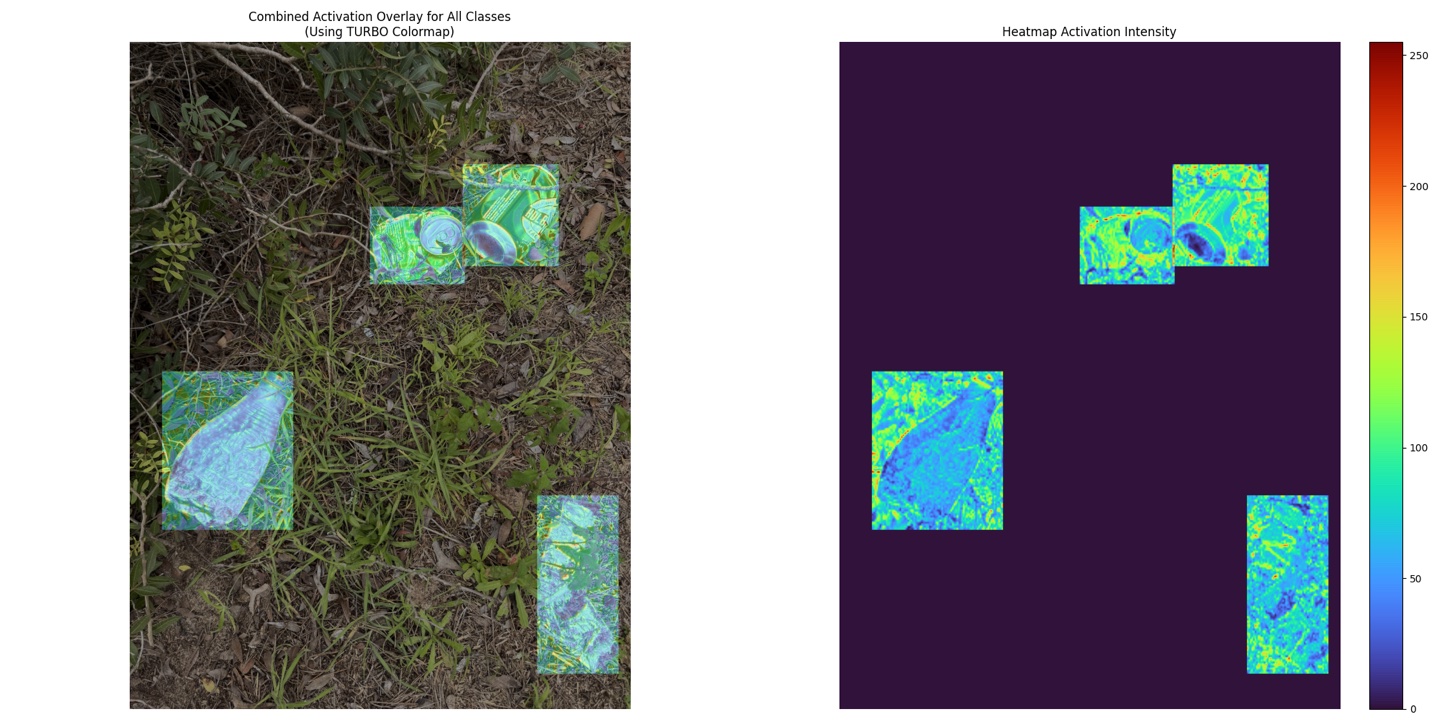}
    \caption{Example of a CAM-based overlay showing the spatial focus of the model during object detection. This image was generated as part of the AIJIM pipeline and helps users visually interpret detection relevance. Refer to Appendix~\ref{app:cam_example} for a detailed explanation.}
    \label{app:cam_example}
\end{figure}

\begin{figure}[htbp]
    \centering
    \includegraphics[width=1.0\columnwidth]{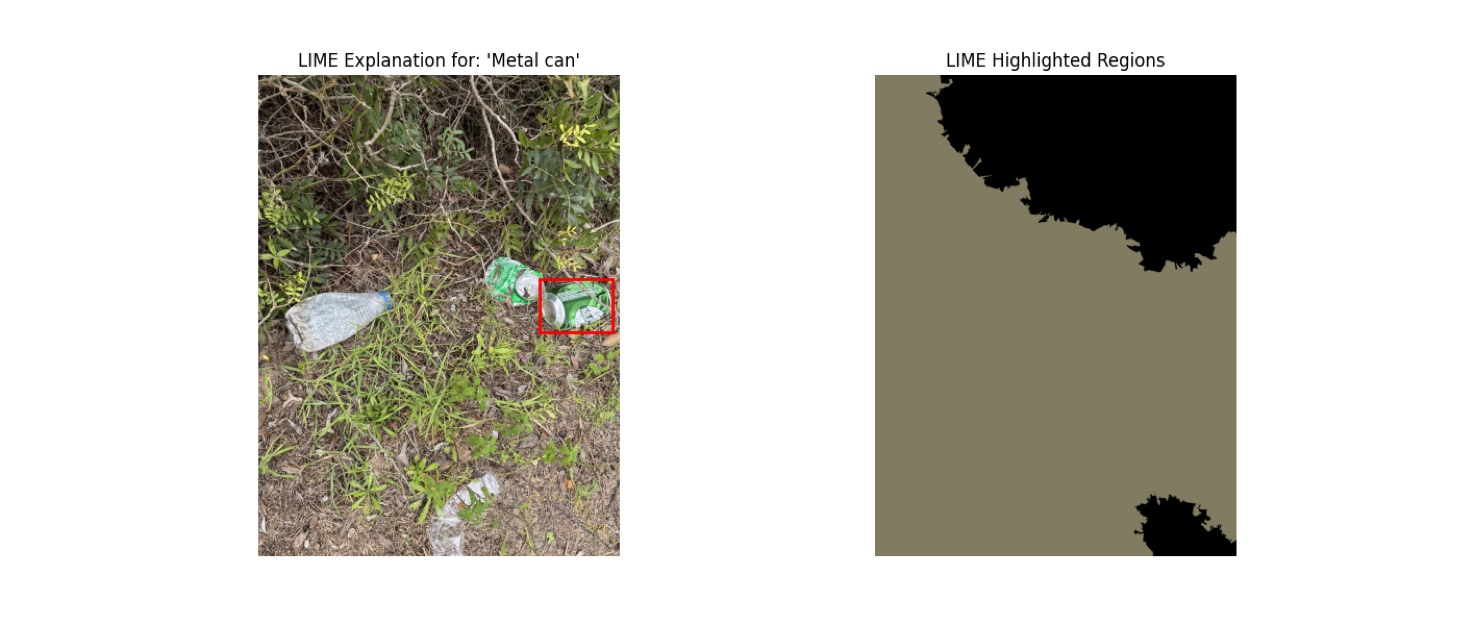}
    \caption{LIME Explanation for Metal Can Detection: Highlighting the key regions contributing to the model's decision. A detailed example can be found in Appendix~\ref{app:lime_example}.}
    \label{app:lime_example}
\end{figure}